\documentstyle[12pt]{article}
\newcommand{\BEQ}{\begin{equation}}    
\newcommand{\BEA}{\begin{eqnarray}}
\newcommand{\EEQ}{\end{equation}}      
\newcommand{\EEA}{\end{eqnarray}}
\newcommand{\eps}{\epsilon}                      
\newcommand{\zeile}[1]{\vskip #1 \baselineskip}  

\newcommand{\build}[3]{\mathrel{\mathop{\kern 0pt#1}\limits_{#2}^{#3} }}

\catcode`\@=11
\def\numberbysection{\@addtoreset{equation}{section}
        \def\theequation{\thesection.\arabic{equation}}}

\begin{document}
\baselineskip 0.3in
%
%
\begin{titlepage}
\null
\vskip 1cm
\begin{center}
\vskip 0.5in
{\Large \bf Finite-size scaling and universality in the spin-1 quantum
XY chain}
\vskip 0.5in
Walter Hofstetter and Malte Henkel\footnote{Address after 1$^{st}$ October
1995: Laboratoire de Physique du Solide, Universit\'e Henri Poincar\'e,
BP 239, F - 54506 Vand{\oe}uvre-l\`es-Nancy Cedex, France}
 \\[.3in]
{\em Theoretical Physics, Department of Physics, \\
University of Oxford, 1 Keble Road, Oxford OX1 3NP, UK}
\end{center}
\zeile{2}
%
\begin{abstract}
The spin-1 XY chain in a transverse field is studied
using finite-size scaling. The ground state phase diagram displays a
paramagnetic, an ordered ferromagnetic and an ordered oscillatory
phase. The paramagnetic-ferromagnetic transition line belongs to the
universality class of the $2D$ Ising model. Along this line, universality is
confirmed for the finite-size scaling functions of several correlation lengths
and for the conformal operator content.
\end{abstract}
\zeile{2}
\end{titlepage}

\newpage
%
%
In modern theories of (equilibrium)
critical phenomena, the notions of scaling and
universality play a central role. These notions are particularly useful
when applied to finite systems using finite-size scaling techniques,
see \cite{Priv94} for an
extensive review. In this work,
we study the effects of varying the spin quantum
number on the thermodynamics of the well-known XY quantum chain in a transverse
field. For spin $\frac{1}{2}$, this model is exactly integrable in terms
of free fermions and many of its properties are well studied, see
\cite{Kats62,Burk87}. Besides being of interest in its own right
(i.e. for the influence of the quantum effects on the order parameter
profile \cite{Henk95}), this
quantum Hamiltonian also arises from the master equation description of
several
{\em non-equilibrium} statistical systems, see \cite{Gryn94}.
Here we consider the spin-1 variant of this model,
with the Hamiltonian
\BEQ  \label{hamiltonian}
H = -\frac{1}{\zeta}
\sum_{n=1}^{N} \left[ h S^z_n + \frac{1+\eta}{2} S^x_n S^x_{n+1}
+\frac{1-\eta}{2} S^y_n S^y_{n+1} \right]
\EEQ
where $h$ is the transverse field, $\eta$ measures the spin anisotropy,
$\zeta$ is a normalization constant and
$N$ is the system size. We use periodic boundary conditions. Finally,
the $S^{x,y,z}$ are spin-1 matrices
\BEQ
S^x = \frac{1}{\sqrt{2}} \left( \begin{array}{ccc} 0 & 1 & 0 \\
                                                   1 & 0 & 1 \\
                                                   0 & 1 & 0 \end{array}
\right) \;\; , \;\;
S^y = \frac{1}{\sqrt{2}} \left( \begin{array}{ccc} 0 & -i & 0 \\
                                                   i & 0 & -i \\
                                                   0 & i & 0 \end{array}
\right) \;\; , \;\;
S^z = \left( \begin{array}{ccc} 1 & 0 & 0 \\
                                0 & 0 & 0 \\
                                0 & 0 & -1 \end{array} \right)
\EEQ
(Spin-1 Ising models were recently proposed to describe the adsorption of CO on
graphite, see \cite{Wiec93}.)
We are interested in the ground state energy $E_0$, which plays the role
of the equilibrium free energy (for reviews see \cite{Priv90,Chri93})
and in the
correlation lengths $\xi_i$, related to the
exponential decay of two-point correlation functions,
given by the energy gaps $\xi_i^{-1} = E_i - E_0$. We calculate the
low-lying spectrum of $H$ for finite $N$
(up to $N=14$) using the Lanczos algorithm and
then extrapolate towards $N\to\infty$, see \cite{Chri93} for details.
The quantum Hamiltonian $H$ commutes
with the charge operator $Q$, the parity operator $P$ and the translation
operator $T$ defined by
\begin{equation}
Q=\prod_{i=1}^{N}\left(2(S_i^z)^2-1\right) \;\; , \;\;\;
P S^{x,y,z}_n P^{\dagger} = S^{x,y,z}_{N+1-n} \;\; , \;\;\;
T S^{x,y,z}_n T^{\dagger} = S^{x,y,z}_{n+1}
\end{equation}
Eigenstates of $H$ are thus characterized by the eigenvalues of $Q,P$ and $T$,
which serves to block-diagonalize $H$.

Our first task is to determine the phase diagram, shown in Fig.~1. We recognize
three distinct phases. The first transition, between the paramagnetic phase
P and the ferromagnetic phase~F, is found from conventional finite-size
scaling and will be shown below to be in the $2D$ Ising universality class.
Close to a conventional
critical point of second order, the following finite-size scaling
form for the inverse correlation lengths is expected \cite{Priv84,Priv94}
\BEQ   \label{scaling}
\xi_i^{-1} = N^{-1} S_i \left( C N^y (h-h_c) \right)
\EEQ
where $h_c = h_c(\eta)$ is the critical point,
$y=2-x_{\eps}$ a critical exponent,
$C$ is a non-universal metric factor and
$S_i$ is a {\em universal}\  scaling function.
In particular, from $2D$ conformal invariance, it follows
that $S_i(0) = 2\pi x_i$ \cite{Card84}, where $x_i$ is a universal
critical exponent. Now, the critical point $h_c$ can be found from
phenomenological renormalization \cite{Priv90}. The results, extrapolated
to $N\to\infty$, are displayed
in table~\ref{tab1}. For $\eta=1$ we find agreement with the earlier result
\cite{Gehl94} $h_c \simeq 1.3259$.
\begin{table} \begin{center}
\begin{tabular}{|c|ccccccc|} \hline
$\eta$ & 0.05 & 0.1 & 0.15 & 0.3 & 0.5 & 0.7 & 1.0 \\ \hline
$h_c$  & 1.002(1) & 1.011(1) & 1.0210(1) & 1.0637(1) & 1.1325(1) &
1.2080(1) & 1.32587(1) \\
$\zeta$ & -- & 0.170(1) & 0.239(1) & 0.4252(1) & 0.6416(1)
& 0.8417(1) & 1.12706(1) \\ \hline
\end{tabular}
\caption[Critical points]{Critical points $h_c(\eta)$ and conformal
normalization $\zeta(\eta)$ for the spin-1 XY model along the Ising
line. The numbers in brackets give the estimated uncertainty in the last digit.
\label{tab1}} \end{center}
\end{table}

The second transition occurs between the ferromagnetic phase F and a new
`oscillatory' phase~O. This transition is well known for the
spin-$\frac{1}{2}$ case \cite{Kats62} and occurs along the line $h=h_o(\eta)$
where\footnote{Along this line $H$ can also
be obtained from the master equation
of certain $1D$ stochastic systems \cite{Gryn94}.}
\begin{equation} \label{OCF}
\eta^2 + h_o(\eta)^2=1
\end{equation}
For spin-$\frac{1}{2}$ it is known that while in the F phase the
connected spin-spin correlation function $<S^x_R S^x_0>_c$ decays
monotonously with $R$, the oscillatory phase is characterized by a
new wave vector $K$ which modulates the spin-spin correlator \cite{Kats62}
\begin{equation}
<S^x_R S^x_0>_c \sim R^{-2} \exp(-2R/\xi ) \cos(KR)
\end{equation}
Furthermore, in the oscillatory phase there
are level crossings in the ground state energy which
occur already for {\em finite} values of the number of
sites $N$ \cite{Hoe85,Henk95}.
It was shown in \cite{Hoe85} that the location
$h_k(N)$ of the $k$-th level crossing satisfies a finite-size scaling law
\begin{equation} \label{zeros}
h_o - h_k(N) \sim N^{-1/\nu}
\end{equation}
where the exponent $\nu$ describes the scaling of the wave vector $K \sim
(h_o-h)^{\nu}$ in the
vicinity of the O/F transition line (for $h\leq h_o$ ).
For spin $S=\frac{1}{2}$, it is known that $\nu=1/2$ \cite{Hoe85}.

We now ask whether a similar transition occurs for larger values of $S$.
Indeed, it is known that for {\em arbitrary} spin $S$ and periodic boundary
conditions, the ground state energy of $H$ is
doubly degenerate at $h=h_o(\eta)$
\cite{Kur82}. For spin $S=1$, we have checked numerically that the
first ground state level crossing
$h_1(N)$ always occurs at $h=h_o(\eta)$ for all finite $N$. In addition,
the second crossing $h_2(N)$ converges towards $h_o(\eta)$, as apparent
from the extrapolated data in table~\ref{tab2}. The exponent found from
eq.~(\ref{zeros}) is consistent with $\nu\simeq 1/2$,
independently of $\eta$ and in agreement with the exact result
for spin-$\frac{1}{2}$. This supports universality along the F/O transition
line. In fact, having confirmed the same finite-size scaling behaviour of the
level crossings in the ground state energy for both spin $S=\frac{1}{2}$ and
$S=1$, we expect the
features of the oscillatory phase known \cite{Kats62,Hoe85}
from $S=\frac{1}{2}$ to be present for $S=1$ as well.
\begin{table}  \begin{center}
\begin{tabular}{|c|cccc|}    \hline
$\eta$ & 0.1 & 0.3 & 0.5 & 0.7  \\ \hline
$h_o$ & 0.9955(3) & 0.9538(3) & 0.8657(3) & 0.7136(5)  \\
$\nu$ & 0.50(1) & 0.48(2) & 0.48(2) & 0.47(3) \\ \hline
\end{tabular}
\caption[Extrapolated finite-size estimates]{Extrapolated finite-size
estimates for the critical point $h_o$ and
the exponent $\nu$ for the spin-1 XY model (\ref{hamiltonian}) as determined
from the second ground state level crossing $h_2(N)$ of eq.~(\ref{zeros}).
\label{tab2} }
\end{center} \end{table}

{}From now on, we concentrate on the P/F transition line. We expect this
transition to be in the $2D$ Ising universality class, if $\eta\neq 0$.
To see this, we compare the
low-lying excitation spectrum of $H$ with the prediction of conformal
invariance \cite{Card87,Chri93},
following the steps outlined for the spin-$\frac{1}{2}$ case
in \cite[p. 135]{Chri93}.
Conformal field theory states that, after subtraction of a
purely extensive term, $H$ can be written in the form
\begin{equation}
H=\frac{2\pi}{N} (L_0+\bar L_0)-\frac{\pi c}{6N} + o\left( \frac{1}{N}
\right)
\end{equation}
where $c$ is the central charge and $L_0$, $\bar L_0$ are generators
of the Virasoro algebra which acts as a dynamical symmetry for $H$.
As a consequence, eigenstates can be grouped into `conformal towers',
each represented by exactly one primary operator with conformal
weights~$(\Delta ,\bar \Delta )$. The scaling dimension of the corresponding
eigenstate is $x=\Delta + \bar \Delta$. The scaled energies and
momenta take the form
\begin{equation}
{\cal E}_{\Delta, \bar \Delta }(I,\bar I) \equiv \lim_{N \to \infty}
\left( E_{\Delta ,\bar \Delta }(I,\bar I) - E_0 \right)\cdot \frac{N}{2\pi} =
(\Delta +I)+(\bar \Delta + \bar I)
\end{equation}
\begin{equation}
{\cal P}_{\Delta ,\bar\Delta }(I,\bar I) \equiv \lim_{N \to \infty}
P_{\Delta ,\bar \Delta}(I, \bar I)\cdot \frac{N}{2\pi}=
(\Delta +I)-(\bar \Delta + \bar I)
\end{equation}
with $I$, $\bar I$ integer.
$E_{\Delta,\bar\Delta},P_{\Delta,\bar\Delta}$, respectively,
are the eigenvalues of $H,i \ln T$ and $E_0$ is the ground state
energy. However, the application of these relations requires that the
scaled energies $\cal E$ and momenta $\cal P$ are measured in the same
units, thus
fixing the normalization $\zeta$ of $H$ accordingly. We find $\zeta$
by demanding that ${\cal E}_{0,0}(2,0)=2$ throughout \cite{Card87}.
The results for $\zeta$ are given in table~\ref{tab1}.

Next, we determine the central charge. For $\eta=1$, we find
$c=0.49999(1)$, close to the expected $c=\frac{1}{2}$ for the
$2D$ Ising universality class.
We did not compute $c$ explicitly for other values of $\eta$, but expect
$c$ to be $\eta$-independent. In order to check
the complete operator content, we give the
extrapolated values of the scaled energies in the charge sectors $Q=0$
and $Q=1$ in tables \ref{tab3} and~\ref{tab4}. When comparing these
spectra to the expected operator content of the $2D$
Ising model \cite{Card87,Chri93}, namely
for the $Q=0$ sector the conformal towers generated by the primary
operators ($0,0$) and ($\frac{1}{2}$,$\frac{1}{2}$)
(which correspond to the vacuum {\bf 1} and the energy density $\eps$)
and for the $Q=1$ sector the conformal tower generated by
($\frac{1}{16}$,$\frac{1}{16}$) (which corresponds to the order parameter
density $\sigma$), we find complete agreement. In particular, we read off
the scaling dimensions $x_{\sigma}=\frac{1}{8}$ and $x_{\eps}=1$ which
determine the bulk critical exponents.

\begin{table}  \begin{center}
\begin{tabular}{|c|c|c|c|c|c|} \hline
5 	&         &   ?     &  5.03(3)  &   ?      &  4.8(2)     	\\
  	&         &   ?     &    ?       &   ?      &     ?
	\\    \hline
4 	&   ?      & 3.98   & -         & 4.00(1) & 3.95(3), 3.94(3)   	\\
  	&   ?      & 3.99(2) & -         & 3.9(1)  &    ?
	  	\\  \hline
3 	& 3.00(1) &  -      &  3.000(1) & 2.999(1) & -     		\\
  	& 3.00(1) & -      & 3.001(1)   & 3.00(2)  &  -
	\\   \hline
2 	& -       & 2.00000(1) & 2       & -       & -         	\\
  	& -      & 2.000(1)   & 2           & -          & -
	\\  \hline
1 	& 1.00000(1) & -     & -          & -        & -        	\\
  	& 1.0002(2) &  -     &  -         & -        & -       	\\   \hline
${\cal E}=0$ & 0     &  -     &   -         &   -       & -     	\\
  	& 0        &  -     &   -        &   -        & -      	\\ \hline \hline
  	& ${\cal P}=0$     &  1     &  2          &  3        & 4
	\\ \hline
\end{tabular}
\caption[Spectrum for $Q=0$.]{Low lying excitations for
charge $Q=0$ at the critical
point. In each box, the upper value corresponds to $\eta = 1$,
the lower one to $\eta = 0.3$.
A dash indicates that no level is present, a `?' indicates that the finite-size
data did not converge. For ${\cal P}=0$, all eigenstates shown have parity
$P=+1$ and the lowest excitations with $P=-1$ occur for ${\cal E}\geq 6$.}
\label{tab3} \end{center}
\end{table}

\begin{table}  \begin{center}
\begin{tabular}{|c|c|c|c|c|}  \hline
$5\frac{1}{8}$	&  ?        & 5.2(1), 5.18(2)   &  ?        &   ?      	\\
		&  ?        &    ?      &    ?      &       ?
	\\  \hline
$4\frac{1}{8}$	& 4.123(2) & -        & 4.128(2), 4.13(1) &  -       	\\
		& 4.1(1)   & -        & 4.13(1), ?  &  -
		\\ \hline
$3\frac{1}{8}$	& -        & 3.124(1) & -        & 3.125(1), 3.124(1) 	\\
		& -        & 3.12(1)  & -        & 3.1(1), 3.1(1)
	\\ \hline
$2\frac{1}{8}$	& 2.1249(1) & -       & 2.1251(2) & -        	\\
		& 2.126(2) & -        & 2.121(2) & -
		\\ \hline
$1\frac{1}{8}$	& -        & 1.12501(1) &  -     &  -         	\\
		& -        & 1.1249(1) &  -      &  -
		\\ \hline
${\cal E}=\frac{1}{8}$     & 0.12499(1) & -  &  -       & -     	\\
		& 0.1249(1) & -       & -        & -
	\\ \hline \hline
		& ${\cal P}=0$       & 1       &  2       &  3
	\\ \hline
\end{tabular}
\caption{Low lying excitations for charge $Q=1$. In each box, the upper
value corresponds to $\eta = 1$, the lower one to $\eta = 0.3$.}
\label{tab4} \end{center}
\end{table}

{}We now look at the finite-size scaling functions for the spin-spin
and energy-energy correlation lengths $\xi_{\sigma,\eps}^{-1} = N^{-1}
S_{\sigma,\eps}(C_{\sigma,\eps} z)$, see eq.~(\ref{scaling}).
{}From universality with $S=\frac{1}{2}$, we expect \cite{Burk87}
\begin{equation}
\frac{1}{2\pi}S_\sigma ( C_{\sigma} z)
=\frac{1}{8} + \frac{1}{4\pi} C_{\sigma} z
+\frac{\ln 2}{4\pi^2}
\left( C_\sigma z\right)^2 +\frac{1}{2} R_{1\frac{1}{2} ,0}
\left( \frac{{(C_\sigma z)}^2}{4\pi^2} \right)
-\frac{1}{8} R_{1\frac{1}{2} ,0} \left(
\frac{{(C_\sigma z)}^2}{\pi^2} \right)
\end{equation}
\begin{equation}
\frac{1}{2\pi}S_\epsilon (C_{\eps} z)
=\sqrt{ 1+\frac{{(C_\epsilon z)}^2}{\pi^2 }}
\end{equation}
where $z=N(h-h_c)$ is the finite-size scaling variable and
$R_{1\frac{1}{2},0}(x)$
is a remnant function \cite{Fish72}. The spin-dependence should only enter
into the metric
factors $C_{\sigma}$ and $C_{\eps}$ which are determined from $S_{\sigma}$ and
$S_{\eps}$, respectively. In Figure~2, we display the extrapolated finite-size
data of $S_{\sigma,\eps}$ for $\eta=0.7$ and find that they match nicely
with the expected functional form. This confirms universality.
Similar plots are obtained for other values of $\eta$. The results for the
metric factors are collected in table~\ref{tab5}.
\begin{table}  \begin{center}
\begin{tabular}{|c|cccc|}   \hline
$\eta$        &  0.3      &   0.5    &  0.7   &  1.0      \\  \hline
$C_\epsilon$  &  2.86(2)  & 2.01(2)  & 1.57(2) & 1.21(2)   \\
$C_\sigma$    & 2.86(2)   & 1.98(2)  & 1.57(2) & 1.20(2)   \\ \hline
\end{tabular}
\caption{Non-universal metric coefficients $C_\epsilon$, $C_\sigma$ found from
the scaling fucntions $S_{\eps}$ and $S_{\sigma}$, respectively.}
\label{tab5} \end{center}
\end{table}
Our results are consistent with
\begin{equation} \label{CRes}
C_\sigma (\eta) = C_\epsilon (\eta) = C(\eta)
\end{equation}
It is interesting to compare these with the conformal normalization
$\zeta(\eta)$ from table~\ref{tab1}. Our data are roughly consistent
with a linear relation
$C^{-1} (\eta ) = \alpha\cdot\zeta (\eta )$ with $\alpha \simeq 0.75$.

A few comments are in order. Firstly, the observation eq.~(\ref{CRes})
that the numerical value of the metric factor is independent of the
physical quantity used for its determination, is certainly
in agreement with the scaling expectation eq.~(\ref{scaling}) \cite{Priv94}.
Similar results were recently reported for $2D$
percolation \cite{Hu95}, where it was
also checked that the metric factors are independent of the boundary
conditions. Secondly, our results confirm earlier work \cite{Debi87}
on the universality of the finite-size scaling function $S_{\sigma}$ in
the $2D$ spin-1 Ising model. Thirdly, the observed linear relation
between the conformal normalization $\zeta$ and the metric factor $C$ can
be understood in terms of conformal perturbation theory, see
\cite{Card87,Chri93}. In that framework,
one would write for the non-critical quantum Hamiltonian
$H = \frac{1}{\zeta}(H_c + g \phi)$, where $H_c$ is the critical point quantum
Hamiltonian, $\phi$ a perturbing relevant operator and $g$ a non-universal
coupling. In our case, $\phi = \eps$, the energy density and $g = h-h_c$.
Since a given quantum Hamiltonian must in general be normalized to make
conformal invariance applicable (see above), we note that into a perturbative
calculation of the energy spectrum only the finite-size scaling variable
$N^{2-x_{\phi}} (h-h_c)/\zeta$ enters. That is consistent with our finding
$C \sim \zeta^{-1}$.

In conclusion, we have investigated the ground state
phase diagram of the spin-1
quantum XY chain in a transverse magnetic field. The structure of
the phase diagram, obtained from finite-size scaling,
is found to be very similar to the known
spin-$\frac{1}{2}$ case. We have explicitly confirmed the universality
of the Ising line, with respect to the spin $S$ as well as
the spin anisotropy $\eta$, considering both the conformal
operator content and
the finite-size scaling functions of the first two correlation lengths. \\

\noindent{\bf Acknowlegdements:} WH would like to thank the Subdepartment
of Theoretical Physics for hospitality and the Stiftung Maximilianeum,
the Bayerische Begabtenf\"orderung and the Studienstiftung des Deutschen
Volkes for financial support. MH was supported by a grant of the EC
`Human Capital and Mobility' programme.

\zeile{5}
\begin{center}
{\Large \bf Figure captions}
\end{center}
\noindent {\bf Figure 1} Ground state phase diagram of the spin-1
XY model (\ref{hamiltonian}). P labels the disordered paramagnetic phase,
F labels the ordered ferromagnetic phase and O labels the ordered oscillatory
phase. The dotted line gives the P/F transition which for $\eta\neq 0$ is in
the $2D$ Ising universality class. The dashed line, given by
$\eta = 2 \sqrt{ (h-1)/5}$ is the approximation to the P/F line as found
from second  order perturbation theory around $\eta=0$. The full line
represents the F/O transition as given by (\ref{OCF}). \\
\noindent {\bf Figure 2} Finite-size scaling
functions $S_{\sigma}(C_{\sigma} z)$ (lower curve) and
$S_{\eps}(C_{\eps} z)$ (upper curve) as a
function of the finite-size scaling variable $z=N(h-h_c)$ for $\eta=0.7$
as compared to the extrapolated finite-lattice estimates (points). \\

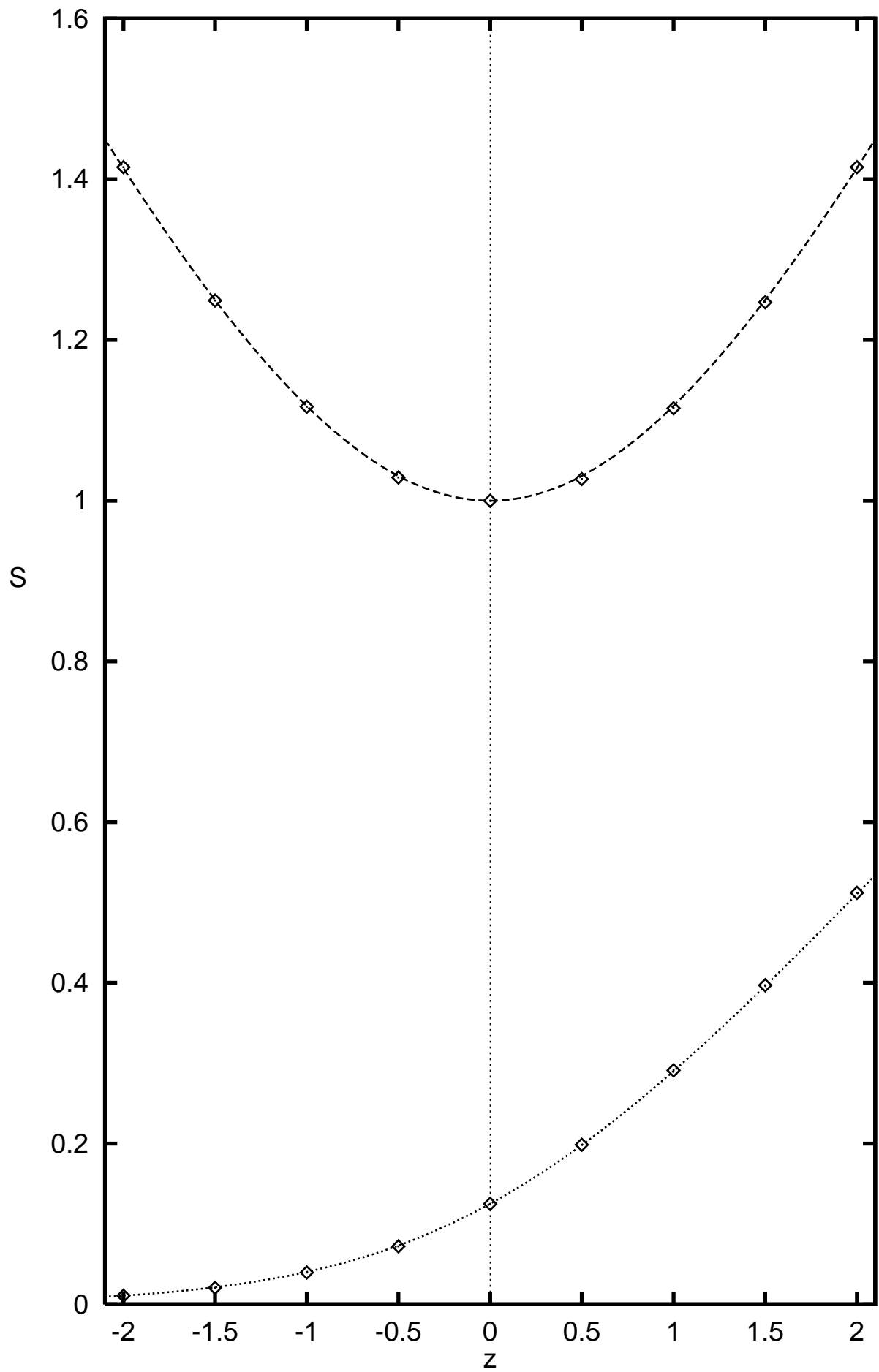
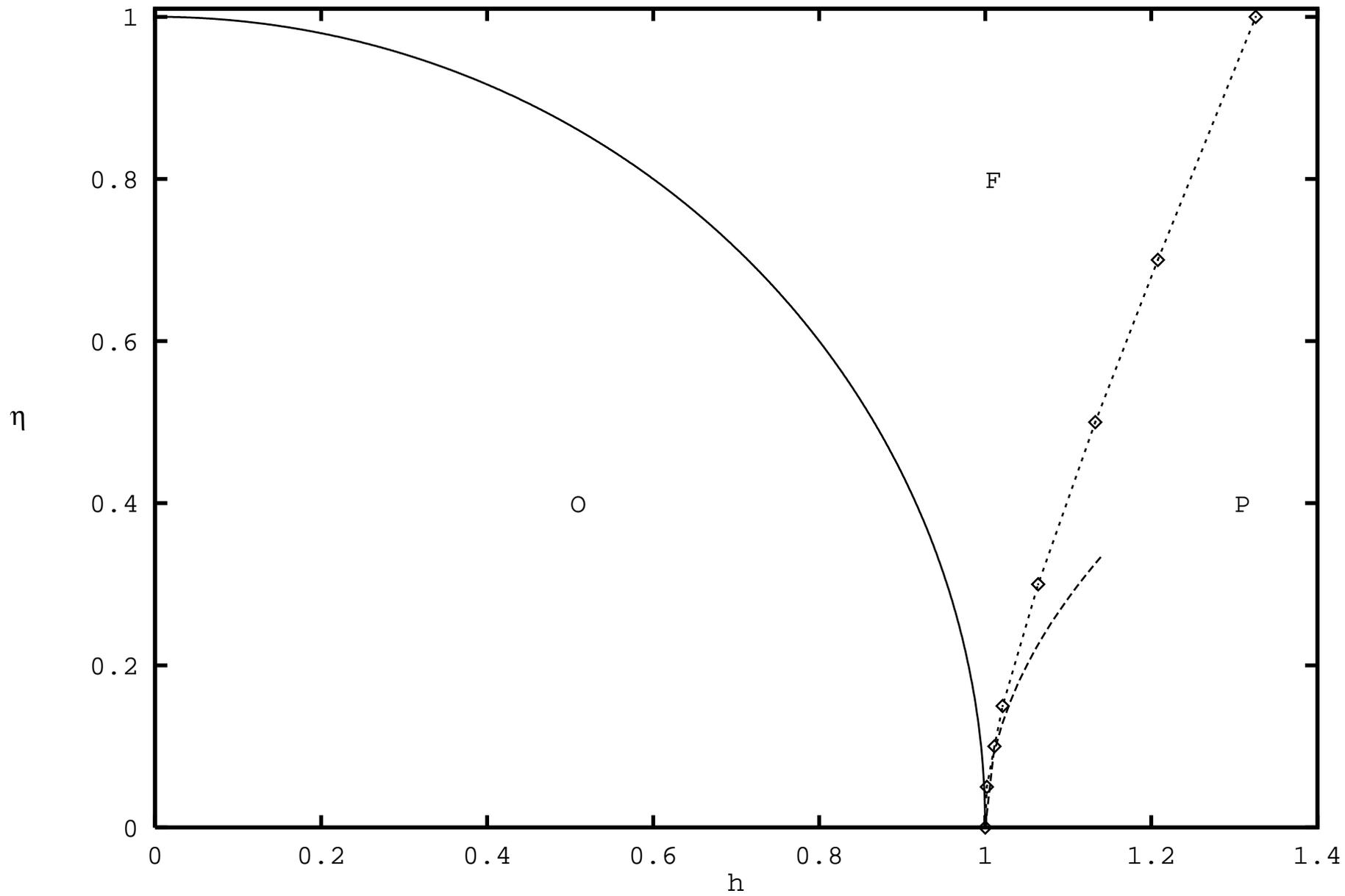
\end{document}